\documentclass[a4paper,showkeys,nofootinbib,aps]{revtex4}
\usepackage{amsmath}
\usepackage{epsfig}
\begin{document}
\title{Azimuthal Asymmetry and Ratio $R= F_L / F_T$\\ 
as Probes of the Charm Content of the Proton\\}
\author{N.Ya.~Ivanov}
\affiliation{Yerevan Physics Institute, Alikhanian Br.2, 0036 Yerevan, Armenia}
\begin{abstract}
\noindent We study two experimental ways to measure the heavy-quark content of the proton: 
using the Callan-Gross ratio $R(x,Q^2)=F_L/F_T$ and/or azimuthal $\cos(2\varphi)$ asymmetry in deep inelastic lepton-nucleon scattering. Our approach is based on the perturbative stability of the QCD predictions for these two quantities. We resume the mass logarithms of the type $\alpha_{s}\ln\left( Q^{2}/m^{2}\right)$ and conclude that heavy-quark densities in the nucleon can, in principle, be determined from data on the Callan–Gross ratio and/or azimuthal asymmetry. In particular, the charm content of the proton can be measured in
future studies at the proposed Large Hadron–Electron (LHeC) and Electron–Ion (EIC) Colliders.
\end{abstract}
\keywords{Perturbative QCD, Heavy-Quark Leptoproduction, Charm Content, Mass Logarithms Resummation, Callan-Gross Ratio, Azimuthal Asymmetry}
\maketitle

\section{Introduction}

The notion of the intrinsic charm (IC) content of the proton has been introduced about 30 years ago in works \cite{BHPS}. It was shown that, in the light-cone Fock space picture, it is natural to expect a five-quark state contribution of the type $\left\vert uudc\bar{c}\right\rangle$ to the proton wave function \cite{brod1}. This component can be generated by
$gg\rightarrow c\bar{c}$ fluctuations inside the proton where the gluons are coupled to different valence quarks. The original concept of the charm density in the proton \cite{BHPS} has nonperturbative nature because the five-quark contribution $\left\vert uudc\bar{c}\right\rangle$ scales as $1/m^{2}$ where $m$ is the $c$-quark mass \cite{polyakov}.

In the middle of nineties, another point of view on the charm content of the proton has been proposed in the framework of the variable-flavor-number scheme (VFNS) of quantum chromodynamics (QCD) \cite{ACOT,collins}.  The VFNS is an approach alternative to the traditional fixed-flavor-number scheme (FFNS) of QCD where only light degrees of
freedom ($u,d,s$ and $g$) are considered as active. Within the VFNS, the mass logarithms of the type $\alpha_{s}\ln\left( Q^{2}/m^{2}\right)$ are resummed through the all orders in perturbation theory into a heavy quark density which evolves with $Q^{2}$ according to the standard DGLAP \cite{grib-lip} evolution equation. Hence this approach introduces the parton distribution functions (PDFs) for the heavy quarks and changes the number of active flavors by one unit when a heavy quark threshold is crossed. Note also that the charm density arises within the VFNS perturbatively via the $g\rightarrow c\bar{c}$ evolution. Some recent developments concerning the VFNS are presented in \cite{chi,SACOT,Thorne-NNLO}. So, the VFNS was introduced to resum the mass logarithms and to improve thus the convergence of perturbative QCD series.

At the present time, both nonperturbative and perturbative charm densities are widely used for a phenomenological description of available data. (A detailed review of the theory and experimental constraints on the charm quark distribution may be found in \cite{pumplin}). In particular, practically all the recent versions of the CTEQ \cite{CTEQ4,CTEQ5,CTEQ6} and MRST \cite{MRST2004} sets of PDFs are based on the VFN schemes and contain a charm density. At the same time, the key question remains open: How to measure the charm content of the proton? The basic theoretical problem is that radiative corrections to the heavy-flavor production cross sections are large: they increase the leading order (LO) results by approximately a factor of two. Moreover, soft-gluon resummation of the threshold Sudakov logarithms indicates that higher-order contributions can also be substantial (for reviews, see \cite{Laenen-Moch,kid1}.) On the other hand, perturbative instability leads to a high 
sensitivity of the theoretical calculations to standard uncertainties in the input QCD parameters: the heavy-quark mass, $m$, factorization and renormalization scales, $\mu _{F}$ and $\mu _{R}$, $\Lambda_{\mathrm{QCD}}$ and PDFs. For this reason, one can only estimate the order of magnitude of the pQCD predictions for charm production cross sections in 
the entire energy range from the fixed-target experiments \cite{Mangano-N-R} to the RHIC collider \cite{R-Vogt}.

Since production cross sections are not perturbatively stable within QCD, they cannot be a good probe of the charm density in the proton. For this reason, it is of special interest 
to study those observables that are well-defined in QCD. Nontrivial examples of such observables were proposed in \cite{we1,we2,we3,we4,we5,we6,we7}, where the azimuthal 
$\cos(2\varphi)$ asymmetry and Callan-Gross ratio $R(x,Q^2)=F_L/F_T$ in heavy quark leptoproduction were analyzed (note also the paper \cite{Almeida-S-V} where the perturbative stability of the QCD predictions for the charge asymmetry in $t$-quark hadroproduction has been observed). It was shown that, contrary to the production cross sections, the azimuthal asymmetry \cite{we2,we4} and the Callan-Gross ratio \cite{we7} in heavy flavor leptoproduction are stable within the FFNS, both parametrically and perturbatively.

\begin{figure}
\begin{center}
\mbox{\epsfig{file=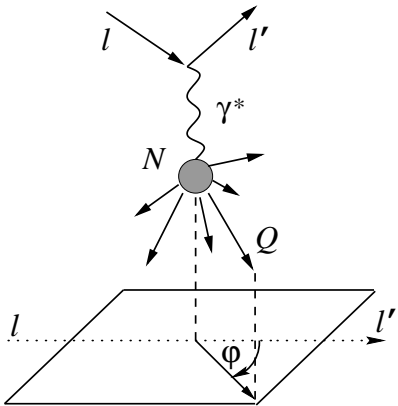,width=150pt}}
\caption{\label{Fig1}\small Definition of the azimuthal
angle $\varphi$ in the target rest frame.}
\end{center}
\end{figure}

In the present paper, we discuss resummation of the mass logarithms of the type $\alpha_{s}\ln\left( Q^{2}/m^{2}\right)$ in heavy quark leptoproduction:
\begin{equation}
l(\ell )+N(p)\rightarrow l(\ell -q)+Q(p_{Q})+X[\overline{Q}](p_{X}). \label{1}
\end{equation}
The cross section of the reaction (\ref{1}) may be written as
\begin{equation}\label{2}
\frac{\mathrm{d}^{3}\sigma_{lN}}{\mathrm{d}x\mathrm{d}Q^{2}\mathrm{d}\varphi}=\frac{2\alpha^{2}_{em}}{Q^4}
\frac{y^2}{1-\epsilon}\left[ F_{T}( x,Q^{2})+ \epsilon F_{L}(x,Q^{2}) 
+ \epsilon F_{A}( x,Q^{2})\cos (2\varphi)+
2\sqrt{\epsilon(1+\epsilon)} F_{I}( x,Q^{2})\cos \varphi\right],
\end{equation}
where $F_{2}(x,Q^2)=2x(F_{T}+F_{L})$, while the quantity $\varepsilon$ measures the degree of longitudinal polarization of the virtual photon in the Breit frame: $\epsilon=2(1-y)\left/\left(1+(1-y)^2\right)\right.$ \cite{dombey}. 
The quantities $x$, $y$, and $Q^2$ are the usual Bjorken kinematic variables while the azimuth $\varphi$ is defined in Fig.~\ref{Fig1}.

In next sections, we will consider resummation of the mass logarithms for the quantities $R(x,Q^2)$ and $A(x,Q^2)$ defined as
\begin{equation}\label{3}
R(x,Q^{2})=\frac{F_{L}}{F_{T}}(x,Q^{2}), \qquad \qquad \qquad \qquad A(x,Q^{2})=2x\frac{F_{A}}{F_{2}}(x,Q^{2}).
\end{equation}

\section{\label{ratio} Resummation of mass logarithms for quantities $F_2$ and $F_L/F_T$ }

\begin{figure}
\begin{center}
\mbox{\epsfig{file=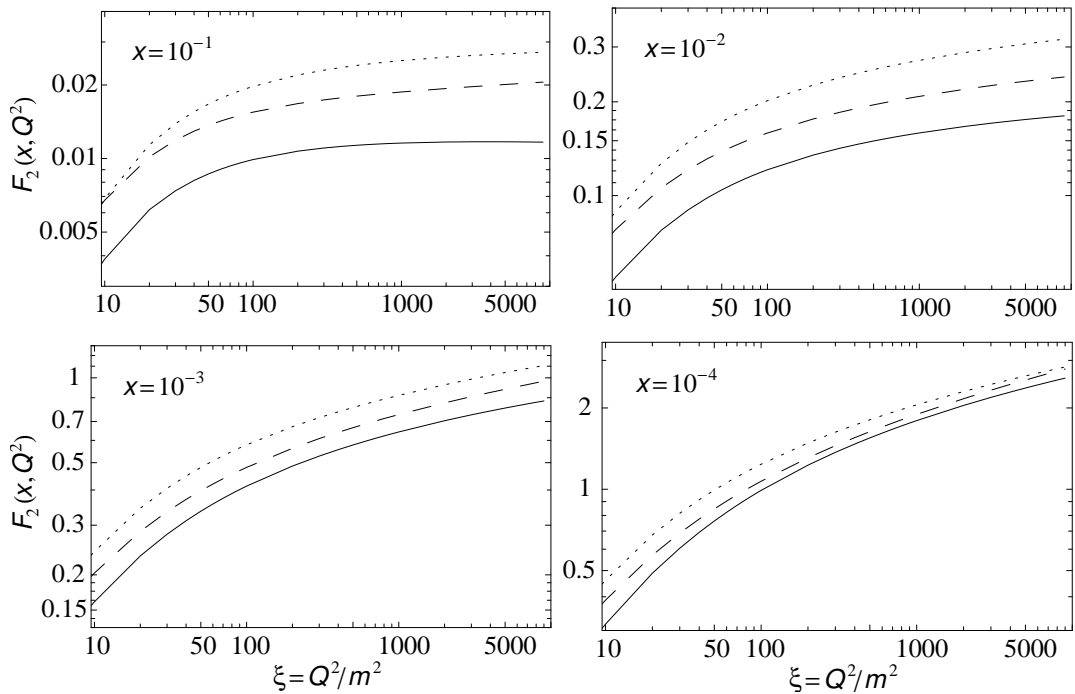,width=400pt}}
\caption{\label{Fig2}\small $Q^2$-dependence of the structure function $F_2(x,Q^2)$ in charm leptoproduction at $x=10^{-1}$, $10^{-2}$, $10^{-3}$, $10^{-4}$. 
Plotted are the LO (solid lines) and NLO (dashed lines) FFNS predictions, as well as the 
ACOT($\chi$) results (dotted curves).}
\end{center}
\end{figure}

\begin{figure}
\begin{center}
\mbox{\epsfig{file=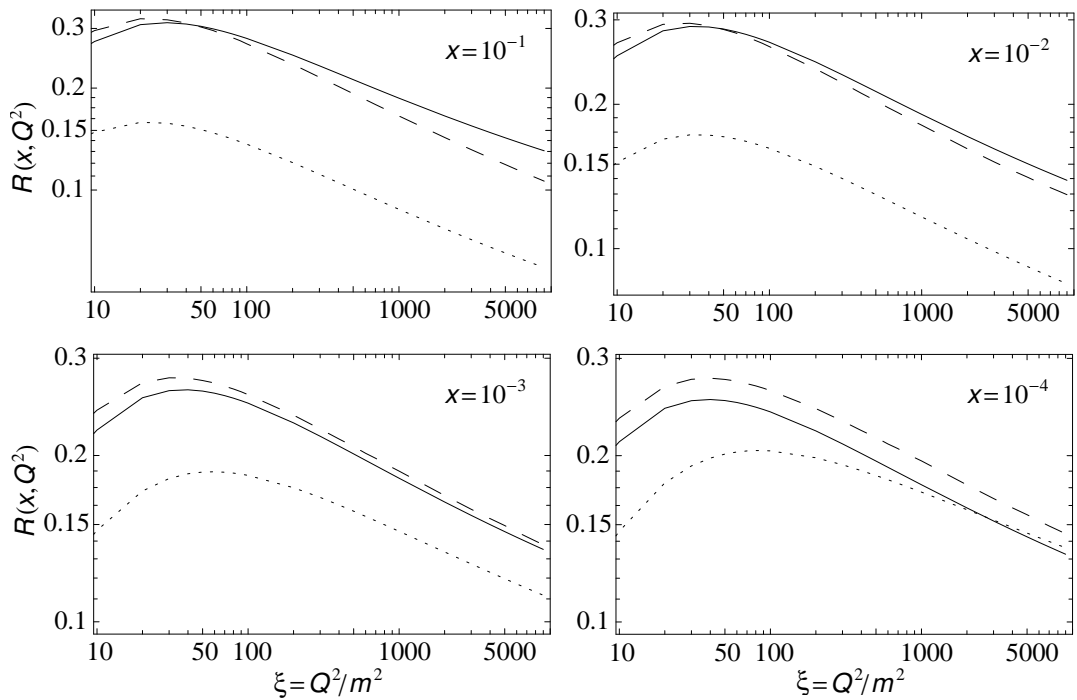,width=400pt}}
\caption{\label{Fig3}\small $Q^2$-dependence of the Callan-Gross ratio $R(x,Q^2)=F_L/F_T$ in charm leptoproduction at $x=10^{-1}$, $10^{-2}$, $10^{-3}$, $10^{-4}$. 
Plotted are the LO (solid lines) and NLO (dashed lines) FFNS predictions, as well as the 
ACOT($\chi$) results (dotted curves).}
\end{center}
\end{figure}

In our analysis, the ACOT($\chi$) VFNS proposed in \cite{chi} is used.
In Figs.~\ref{Fig2} and \ref{Fig3}, we present the LO and next-to-leading order (NLO) 
FFNS predictions for the structure function $F_{2}(x,Q^{2})$ and Callan-Gross ratio $R(x,Q^2)=F_L/F_T$ in charm leptoproduction, and compare them with the corresponding ACOT($\chi$) VFNS results. In our calculations, the CTEQ6M parameterization for PDFs and  $m=1.3$~GeV for $c$-quark mass is used \cite{CTEQ6}. We convolve the NLO CTEQ6M distribution functions with both the LO and NLO partonic cross sections  \cite{Bluemlein} that makes it possible to estimate directly the degree of stability of the FFNS predictions under radiative corrections. The default common value for the factorization and renormalization scales is $\mu=\sqrt{4m_{c}^{2}+Q^{2}}$.

One can see from Fig.~\ref{Fig2} that both the radiative corrections and charm-initiated ACOT($\chi$) contributions to $F_{2}(x,Q^{2})$ are large: they increase the LO FFNS results by approximately a factor of two at $x\sim 10^{-1}$ for all $Q^2$. At the same time, the relative difference between the NLO FFNS and ACOT($\chi$) predictions in not large: it does not exceed $20\%$ for $Q^2/m^2<10^{3}$. We conclude that it will be very difficult to determine the charm content of the proton using only data on $F_2(x,Q^2)$ due to large radiative corrections (with corresponding theoretical uncertainties) to this quantity.

Considering the corresponding predictions for the ratio $R(x,Q^2)$ presented in Fig.~\ref{Fig3}, we see that in this case the NLO and charm-initiated ACOT($\chi$) contributions are strongly different. In particular, the NLO corrections to $R(x,Q^2)$ are small, less than $15\%$, for $x\sim 10^{-3}$--$10^{-1}$ and $Q^2/m^2<10^{4}$. This implies that large radiative contributions to the structure functions $F_T$ and $F_L$ cancel each other in the ratio $F_L/F_T$ with a good accuracy.

On the other hand, the charm content of the proton is predicted to be sizeable within the ACOT($\chi$) scheme which decreases the LO FFNS results by about $50\%$ practically for all values of $Q^2/m^2>10$. The reason for decreasing of $R(x,Q^2)$ within the VFNS is that resummation of the mass logarithms has different effects on the structure functions $F_{T}(x,Q^{2})$ and $F_{L}(x,Q^{2})$. In fact, contrary to the transverse structure function, $F_{T}(x,Q^{2})$, the longitudinal one, $F_{L}(x,Q^{2})$, does not contain leading mass logarithms of the type $\alpha_{s}\ln\left( Q^{2}/m^{2}\right)$ at both LO and NLO \cite{LRSN,BMSMN}. For this reason, resummation of the $\alpha_{s}\ln\left( Q^{2}/m^{2}\right)$ logarithms within the ACOT($\chi$) scheme leads to increasing of the quantity $F_{T}(x,Q^{2})$ but does not affect the function $F_{L}(x,Q^{2})$. We conclude that, contrary to the the production cross sections, the Callan-Gross ratio $R(x,Q^2)=F_L/F_T$ in deep-inelastic leptoproduction of $c$-quark could be a good probe of the charm density in 
the proton.

Note that this observation is practically independent of the PDFs we use. We have verified that the CTEQ versions \cite{CTEQ4,CTEQ5,CTEQ6} of the PDFs lead to sizeable reduction of the LO FFNS predictions for the ratio $R(x,Q^2)$.

As to the low $x<10^{-4}$ behavior of the Callan-Gross ratio, this problem requires resummation of the BFKL \cite{BFKL1} logarithms of the type $\ln (1/x)$  and will be considered in a forthcoming publication.

\section{\label{asymmetry} Resummation of mass logarithms for Azimuthal Asymmetry}

\begin{figure}
\begin{center}
\mbox{\epsfig{file=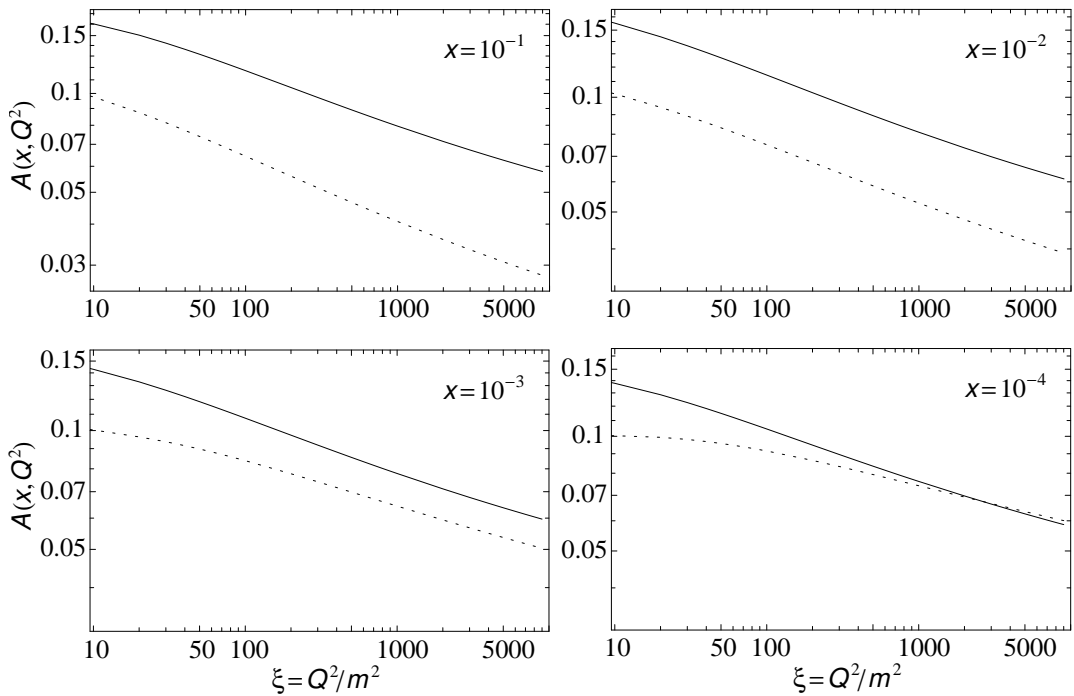,width=400pt}}
\caption{\label{Fig4}\small $Q^2$-dependence of the azimuthal asymmetry $A(x,Q^{2})=2xF_{A}/F_{2}$ in charm leptoproduction at $x=10^{-1}$, $10^{-2}$, $10^{-3}$, $10^{-4}$. 
Plotted are the LO FFNS predictions (solid lines), as well as the ACOT($\chi$) results  (dotted curves).}
\end{center}
\end{figure}

Fig.~\ref{Fig4} shows the ACOT($\chi$) predictions for the azimuthal asymmetry $A(x,Q^2)=2xF_{A}/F_{2}$ at the following values of variable $x=10^{-1}$, $10^{-2}$, $10^{-3}$, $10^{-4}$. For comparison, we plot also the LO FFNS predictions (solid curves). Again, we use the CTEQ6M parametrization of PDFs, $m_c=1.3$~GeV, and $\mu=\sqrt{4m_{c}^{2}+Q^{2}}$.

One can see from Fig.~\ref{Fig4} the following properties of the azimuthal asymmetry: 
the mass logarithms resummation leads to a sizeable decreasing of the LO FFNS predictions for the $\cos(2\varphi)$-asymmetry. Within the ACOT($\chi$) scheme, the charm-initiated contribution reduces the FFNS results for $A(x,Q^{2})$ by about $(30$--$40)\%$ at $x\sim 10^{-2}$--$10^{-1}$. The origin of this reduction is the same as in the case of  $R(x,Q^2)$: 
contrary to $F_{2}$, the azimuth-dependent structure function $F_{A}$ is safe in the limit $m^2\to 0$ at least at LO.

Presently, the exact NLO predictions for the azimuth-dependent structure function $F_{A}$ are not available. However, in \cite{we4} the NLO corrections to the $\cos(2\varphi)$-asymmetry have been estimated within the so-called soft-gluon approximation at $Q^2 \lesssim m^2$. (Note that soft-gluon approximation is unreliable for high $Q^2\gg m^2$). 
It was demonstrated that soft-gluon corrections to both $F_{A}$ and $F_{2}$ are large but cancel each other in their ratio $A(x,Q^{2})=2xF_{A}/F_{2}$ with a good accuracy. 
For this reason, it is natural to expect that the azimuthal $\cos(2\varphi)$-asymmetry is also a perturbatively stable quantity in wide kinematic range of variables $x$ and $Q^{2}$ within the FFNS.

We have also analyzed how the VFNS predictions for $A(x,Q^{2})$ depend on the choice of subtraction prescription. In particular, the schemes proposed in \cite{KS,SACOT} have been considered. We have found that, sufficiently above the production threshold, these subtraction prescriptions also reduce the LO FFNS results for the asymmetry by approximately $(30$--$50)\%$.

Thus, it is shown that impact of the  mass logarithms resummation on the $\cos(2\varphi)$-asymmetry is essential at $x\sim 10^{-2}$--$10^{-1}$ and therefore can be tested experimentally.

\section{\label{conclusion} Conclusion}

In the present paper, we compare the structure function $F_{2}$, Callan-Gross ratio $R=F_L/F_T$, and azimuthal asymmetry $A=2xF_{A}/F_{2}$ in charm leptoproduction as probes of the charm content of the proton. To estimate the charm-initiated contributions, we used the ACOT($\chi$) VFNS \cite{chi} and several CTEQ sets  of PDFs \cite{CTEQ4,CTEQ5,CTEQ6}. 
Our analysis of the radiative and charm-initiated corrections indicates that, in a wide kinematic range, both contributions to the structure function $F_{2}(x,Q^{2})$ have similar $x$ and $Q^2$ behaviors. For this reason, will be difficult to estimate the charm content of the proton using only data on $F_{2}(x,Q^{2})$.

The situation with using the Callan-Gross ratio and azimuthal asymmetry looks more promising. Our analysis shows that resummation of the mass logarithms leads to reduction of the Born  predictions for $R(x,Q^2)$ and $A(x,Q^2)$ by $(30$--$50)\%$ at $x\sim 10^{-2}$--$10^{-1}$ and $Q^2\gg m^2$. Taking into account the perturbative stability of the Callan-Gross ratio and azimuthal asymmetry within the FFNS established in \cite{we4,we7}, we conclude that the charm density in the proton can, in principle, be determined from high-$Q^2$ data on $R=F_L/F_T$ and $A=2xF_{A}/F_{2}$.

Concerning the experimental aspects, the quantities $R(x,Q^{2})$ and $A(x,Q^{2})$ in charm leptoproduction can be measured in future studies at the proposed EIC \cite{eRHIC} and 
LHeC \cite{LHeC} colliders.

\section{Acknowledgments}

We thank S. I. Alekhin and J. Bl$\ddot{\rm u}$mlein for providing us with fast code \cite{Bluemlein} for numerical calculations of the NLO DIS cross sections. The author is grateful to H. Avakian, S.J. Brodsky and C. Weiss for useful discussions. This work was supported in part by the ANSEF 2010 grant PS-2033 and State Committee of Science of RA, grant 11-1C015.


\begin{thebibliography}{99}
\bibitem{BHPS} S.~J.~Brodsky, P.~Hoyer, C.~Peterson, and N.~Sakai,
Phys.\ Lett.\ B {\bf 93}, 451 (1980);
S.~J.~Brodsky, C.~Peterson, and N.~Sakai, Phys.\ Rev.\ D {\bf 23}, 2745 (1981).
\bibitem{brod1} S.~J.~Brodsky, \emph{"Light-front QCD"}, hep-ph/0412101;
S.~J.~Brodsky, Few Body Syst. {\bf 36}, 35 (2005).
\bibitem{polyakov} M.~Franz, V.~Polyakov, and K.~Goeke,
Phys.\ Rev.\ D {\bf 62}, 074024 (2000).
\bibitem{ACOT} M.~A.~G.~Aivazis, J.~C.~Collins, F.~I.~Olness, and W.~-K.~Tung,
Phys.\ Rev.\ D {\bf 50}, 3102 (1994).
\bibitem{collins} J.~C.~Collins, Phys.\ Rev.\ D {\bf 58}, 094002 (1998).
\bibitem{grib-lip} V.~N.~Gribov and L.~N.~Lipatov, Sov.\ J.\ Nucl.\ Phys. {\bf 15}, 438 (1972);
Y.~L.~Dokshitzer, Sov.\ Phys.\ JETP {\bf 46}, 641 (1977);
G.~Altarelli and G.~Parisi, Nucl.\ Phys.\ B {\bf 126}, 298 (1977).
\bibitem{chi} W.~-K. Tung, S.~Kretzer, and C.~Schmidt, J.\ Phys.\ G {\bf 28}, 983 (2002).
\bibitem{SACOT} M.~Kramer, F.~I.~Olness, and D.~E.~Soper, Phys.\ Rev.\ D {\bf 62}, 096007 (2000).
\bibitem{Thorne-NNLO} R.~S.~Thorne, Phys.\ Rev.\ D {\bf 73}, 054019 (2006);
C.~D.~White and R.~S.~Thorne, Phys.\ Rev.\ D {\bf 74}, 014002 (2006);
W.~K.~Tung, H~.L.~Lai, A.~Belyaev, J.~Pumplin, D.~Stump,
and C.~-P.~Yuan, JHEP {\bf 0702}, 053 (2007);
S.~Kretzer, H.~L.~Lai, F.~I.~Olness and W.~-K.~Tung,
Phys.\ Rev.\ D {\bf 69}, 114005 (2004);
R.~S.~Thorne and W.~K.~Tung, arXiv:0809.0714 [hep-ph];
P.~M.~Nadolsky, arXiv:0809.0945 [hep-ph]; 
M.~Guzzi, P.~M.~Nadolsky, H.~L.~ Lai, and  C.-P.~Yuan, arXiv:1108.5112[hep-ph].
\bibitem{pumplin} J.~Pumplin, Phys.\ Rev.\ D {\bf 73}, 114015 (2006);
S.~J.~Brodsky, B.~Kopeliovich, I.~Schmidt, and J.~Soffer,
Phys.\ Rev.\ D {\bf 73}, 113005 (2006);
J.~Pumplin, H.~L.~Lai, and W.~K.~Tung, Phys.\ Rev.\ D {\bf 75}, 054029 (2007).
\bibitem{CTEQ4}
H.L. Lai, J. Huston, S. Kuhlmann, F. Olness, J. Owens, D. Soper, W.K. Tung and H. Weerts,
Phys.\ Rev.\  D {\bf 55}, 1280 (1997).
\bibitem{CTEQ5} H.L. Lai, J. Huston, S. Kuhlmann, J. Morfin, F. Olness, J.F. Owens,
J. Pumplin and W.K. Tung, Eur.\ Phys.\ J.\ C {\bf 12}, 375 (2000).
\bibitem{CTEQ6} J.~Pumplin, D.~R.~Stump, J.~Huston, H.~L.~Lai, P.~Nadolsky,
and W.~K.~Tung, JHEP {\bf 0207}, 012 (2002).
\bibitem{MRST2004} A.~D.~Martin, R.~G.~Roberts, W.~J.~Stirling, and R.~S.~Thorne,
Phys.\ Lett.\ B {\bf 604}, 61 (2004).
\bibitem{Laenen-Moch} E.~Laenen and S.~-O.~Moch, Phys.\ Rev.\ D {\bf 59}, 034027 (1999).
\bibitem{kid1} 
N.~Kidonakis, Phys.\ Rev.\ D {\bf 64}, 014009 (2001);
N.~Kidonakis, Phys.\ Rev.\ D {\bf 73}, 034001 (2006).
\bibitem{Mangano-N-R} M.~L.~Mangano, P.~Nason, and G.~Ridolfi, Nucl.\ Phys.\ B {\bf 373},
295 (1992);
S.~Frixione, M.~L.~Mangano, P.~Nason, and G.~Ridolfi,
Nucl.\ Phys.\ B {\bf 412}, 225 (1994).
\bibitem{R-Vogt} R.~Vogt, Eur.\ Phys.\ J.\ ST {\bf 155}, 213 (2008).
\bibitem{we1} N.~Ya.~Ivanov, A.~Capella, and A.~B.~Kaidalov,
Nucl.\ Phys.\ B {\bf 586}, 382 (2000).
\bibitem{we2} N.~Ya.~Ivanov, Nucl.\ Phys.\ B {\bf 615}, 266 (2001).
\bibitem{we3} N.~Ya.~Ivanov, P.~E.~Bosted, K.~Griffioen, and S.~E.~Rock,
Nucl.\ Phys.\ B {\bf 650}, 271 (2003).
\bibitem{we4} N.~Ya.~Ivanov, Nucl.\ Phys.\ B {\bf 666}, 88 (2003).
\bibitem{we5}
L.~N.~Ananikyan and N.~Ya.~Ivanov, Phys.\ Rev.\ D {\bf 75}, 014010 (2007).
\bibitem{we6}
L.~N.~Ananikyan and N.~Ya.~Ivanov, Nucl.\ Phys.\ B {\bf 762}, 256 (2007).
\bibitem{we7} N.~Ya.~Ivanov and B.~A.~Kniehl, Eur.\ Phys.\ J.\ C {\bf 59}, 647 (2009); N.~Ya.~Ivanov, Nucl.\ Phys.\ B {\bf 814}, 142 (2009).
\bibitem{Almeida-S-V} L.~G.~Almeida, G.~Sterman, and W.~Vogelsang, 
Phys.\ Rev.\ D {\bf 78}, 014008 (2008).
\bibitem{dombey} N.~Dombey, Rev.\ Mod.\ Phys.\ {\bf 41}, 236 (1969).
\bibitem{Bluemlein} S. I. Alekhin and J. Bluemlein, Phys.\ Lett.\ B {\bf 594}, 299 (2004).
\bibitem{LRSN} E.~Laenen, S.~Riemersma, J.~Smith, and W.~L.~van Neerven,
Nucl.\ Phys.\ B {\bf 392}, 162 (1993).
\bibitem{BMSMN} M.~Buza, Y.~Matiounine, J.~Smith, R.~Migneron, and W.~L.~van Neerven,
Nucl.\ Phys.\ B {\bf 472}, 611 (1996).
\bibitem{BFKL1} E.~A.~Kuraev, L.~N.~Lipatov, and V.~S.~Fadin,
Sov.\ Phys.\ JETP {\bf 44}, 443 (1976);
I.~I.~Balitzki and L.~N.~Lipatov, Sov.\ J.\ Nucl.\ Phys.\ {\bf 28}, 822 (1978);
L.~N.~Lipatov,
Sov.\ Phys.\ JETP {\bf 63}, 904 (1986).
\bibitem{KS} S.~Kretzer and I.~Schienbein, Phys.\ Rev.\ D {\bf 58}, 094035 (1998).
\bibitem{eRHIC} A.~Deshpande, R.~Milner, R.~Venugopalan, and W.~Vogelsang,
Ann.\ Rev.\ Nucl.\ Part.\ Sci.\ {\bf 55}, 165 (2005).
\bibitem{LHeC} J.~B.~Dainton, M.~Klein, P.~Newman, E.~Perez, and F.~Willeke, 
J.\ Inst.\ {\bf 1}, P10001 (2006).
\end{thebibliography}
\end{document}